# A magneto-mechanical accelerometer based on magnetic tunnel junctions

Andrea Meo[1], Francesca Garescì[2], Victor Lopez-Dominguez[3-4], Davi Rodrigues[1], Eleonora Raimondo[5], Vito Puliafito[1], Pedram Khalili Amiri[4,*], Mario Carpentieri[1,*] Giovanni Finocchio[5,*]

[1] *Department of Electrical and Information Engineering, Politecnico of Bari, 70125 Bari, Italy*

[2] *Department of Engineering, University of Messina, I-98166, Messina, Italy*

[3] *Institute of Advanced Materials (INAM), Universitat Jaume I, Castellon, 12006, Spain*

[4] *Department of Electrical and Computer Engineering, Northwestern University, 2145 Sheridan Road, Evanston, 60208, Illinois, USA*

[5] *Department of Mathematical and Computer Sciences, Physical Sciences and Earth Sciences, University of Messina, I-98166, Messina, Italy*

*Corresponding authors: pedram@northwestern.edu, mario.carpentieri@poliba.it, gfinocchio@unime.it

Accelerometers have widespread applications and are an essential component in many areas such as automotive, consumer electronics and industrial applications. Most commercial accelerometers are based on micro-electromechanical system (MEMS) that are limited in downscaling and power consumption. Spintronics-based accelerometers have been proposed as alternatives, however, current proposals suffer from design limitations that result in reliability issues and high cost. Here we propose spintronic accelerometers with magnetic tunnel junctions (MTJs) as building block, which map accelerations into a measurable voltage across the MTJ terminals. The device exploits elastic and dipolar coupling as a sensing mechanism and the spintronic diode effect for the direct read out of the acceleration. The proposed technology represents a potentially competitive and scalable solution to current capacitive MEMS-based approaches that could lead to a step forward in many of the commercial applications.

## I. INTRODUCTION

Accelerometers are an essential component in many applications such as automotive [1], consumer electronics [2], seismic monitoring [3], human-computer interaction [4] and Industry 4.0 [5] with the emerging internet of things (IoT) [6], internet of everything (IoE) [7] and digital twin [8]. Most commercial accelerometers are based on micro-electromechanical systems (MEMSs) built with heterogeneous integration [9–11], and among them, capacitive MEMS-based accelerometers are the most common. They exploit the change of capacitance between a fixed electrode and a movable one attached to a spring that behaves as a proof mass. The change in the capacitance relates to the acceleration acting on the device and it allows the measure of the acceleration [3,12]. The main limitations of MEMS-based accelerometers are the difficulty of downscaling, lack of electromagnetic radiation hardness, and high power consumption, as it is required to charge the capacitive plates [3,13].

Spintronics is now an extremely vibrant area of research, development and applications [14,15]. The compatibility with complementary metal-oxide semiconductor (CMOS) manufacturing processes [16–19] as well as recent advancements in material and device fabrication have been driving the development of the key building blocks of spintronics, i.e., magnetic tunnel junctions (MTJ) [16,17,20], to be ready for exploitation in commercial applications such as magnetic sensors [17,21,22], memories [15,23], oscillators [24], and others [16,17,20]. The active part of the MTJs is composed of two ferromagnetic layers, pinned and free layers, separated by a thin isolating material. The resistance of an MTJ depends on the relative orientation of the free layer magnetization with respect to the fixed layer one.

Some proposals of spintronics-based accelerometers have been proposed in the recent years [25–29], but still this research direction is at its infancy. Such proposals rely on mechanical stresses acting on the MTJ device itself, may it be a moving component of the MTJ [25,29] or a cantilever closing in to the MTJ [26,27]. However, these approaches can result in a potential rapid degradation of the magnetic properties and tunneling magnetoresistance of the MTJ because of their complex structure, causing reliability issues as well as induce structural damage. Moreover, the fabrication of these devices would require the development of ad hoc processes and techniques not compatible with the current processes implemented for MTJs, resulting in expensive and complex products.

Here, we have proposed a spintronic MEMS accelerometer designed by coupling micromagnetic simulations and elastic dynamics, drawing inspiration from capacitive MEMS-based accelerometers. The device concept comprises two MTJs working as spin torque nano oscillators (STNOs) that are magnetically and elastically coupled. One MTJ is realized on top of a fixed substrate (fixed-MTJ) while the other is deposited on top of a substrate which is free to move (free-MTJ) but elastically connected to the fixed one. To work as MEMS, the two MTJs have to be magnetically coupled throughout the whole range of displacements that the substrates can reach. The spintronic diode effect across the fixed-MTJ is used as a read-out mechanism [20,30]. We first design the fixed-MTJ, considering experimentally measured physical parameters, where the magnetization dynamics is characterized by the excitation of a uniform mode. The free-MTJ is designed in such a way that the rectified voltage in the fixed-MTJ changes linearly as a function of the distance between the free and fixed MTJs. Variations of the rectified voltage can be linked directly to the external acceleration acting on the free-MTJ because the magnetization dynamics is much faster than the elastic one. With experimentally achievable parameters, we have predicted rectification voltages on the order of tens of microvolts. Our results show that the spintronic MEMS devices proposed here can be potentially competitive candidates to replace current capacitive MEMS-based accelerometers owing to their compactness, CMOS-compatibility, and high sensitivity.

The paper is organized as follows. Sec. II presents the working principle of the proposed spintronic MEMS accelerometer. The device design and the numerical modeling approach utilized in this study are presented and discussed in Sec. III. The approach is validated by characterizing the fixed-MTJ in Sec. IV, while the two-MTJ system is modelled and analyzed in Sec. V. Sec. VI is dedicated to the simulations of the behavior of a spintronic MEMS device in a potential operational environment. Finally, Sec. VII summarizes the main conclusions of this study.

## II. WORKING PRINCIPLE

Fig. 1(a) shows the established design of a capacitive MEMS-based accelerometer. A proof mass and the electrodes attached to it are displaced with respect to fixed electrodes upon application of an external acceleration, and the change in differential capacitance ($C_1 - C_2$ in the figure) is used to estimate the acceleration. An MTJ-based accelerometer can be designed in different ways. In particular, a concept similar to the capacitive MEMS-based accelerometer can be exploited by considering a pair of MTJs placed on top of two substrates connected via an elastic element, as shown in Fig. 1(b). One substrate is fixed, while the other substrate is free to move along a fixed direction in response to an external excitation, and we refer to the MTJs mounted on top of each substrate as fixed-MTJ and free-MTJ, respectively. In this design, the two MTJs are coupled elastically through the substrate, and magnetically via their dipolar interactions. In the presence of an acceleration, the relative distance between the MTJs changes and is translated into a change in the stray field configuration between them (red lines). The proposed solution relies on the spintronic diode effect [20,31]. The two MTJs are in the self-oscillation regime driven by a constant applied current, both working as STNO [32,33]. An ac current is applied to the fixed-MTJ in order to precisely control the oscillation frequency thanks to the injection locking phenomenon. The spintronic diode effect measurable in the injection-locked states exhibits very high sensitivity enhanced by the dc bias current [20,31,34]. Therefore, the fixed-MTJ works as an active spin torque diode. In addition, the system is designed in such a way that the oscillatory dynamics of the two MTJs is mutually synchronized via the dynamical stray field. In this configuration, the rectified voltage across the fixed-MTJ depends on the distance between the two MTJs. An alternative solution where a coil is placed in between the MTJs to perform the read out is presented in Note S1 of Supplemental Material [32].

However, the simpler and smaller design, together with the use of the spintronic diode effect, make the solution of Fig. 1(b) more appealing. Moreover, the spintronic accelerometer proposed here senses the acceleration directly as voltage variation across the fixed-MTJ. Once amplified by an external circuit, the signal can be used for the evaluation of the acceleration directly in contrast to state of the art integrable MEMS devices where a circuit to convert the capacitance variation into an output voltage and then an amplification are needed [33–35].

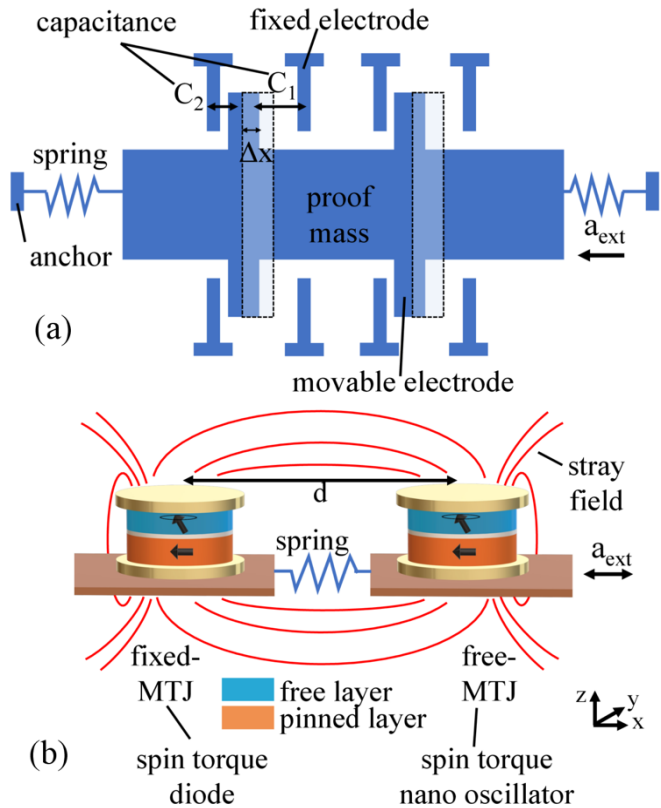

Fig. 1 (a) Schematic description of the working principle of a capacitive MEMS accelerometer composed of fixed electrodes and a movable structure with capacitive "fingers" defined proof mass, connected to anchors via springs. Each finger-electrode $i$ is characterized by a capacitance $C_{i=1,2,...}$. $\Delta x$ is the displacement of the proof mass from the equilibrium position (shaded area) induced by the action of an external acceleration $a_{ext}$. (b) Schematic description of the proposed spintronic MEMS accelerometers. A system of two MTJs, composed of free layer and pinned layer, is coupled via dipolar field (red lines) and elastically via the substates. The free-MTJ, acting as spin torque nano oscillator, is displaced from its equilibrium position (shaded area) of Δx when an external acceleration $a_{\text{ext}}$ is applied, resulting in a change in the MTJs center-to-center distance ($d$) and thus in the stray field distribution. The read out mechanism is based on the spintronic diode effect on the fixed-MTJ that act as spin torque diode. The Cartesian coordinate system used in this paper is also included.

### III. MODELING

The magnetic behavior of the spintronic MEMS accelerometer is simulated by numerically integrating two coupled Landau-Lifshitz-Gilbert-Slonczewski (LLGS) equations for the free- and fixed-MTJ [36–38], whereas the elastic dynamics is described in Eq. 2:

$$\begin{cases} \frac{d\boldsymbol{m_1}}{d\tau} = -\frac{1}{1+\alpha^2}(\boldsymbol{m_1} \times \boldsymbol{h}_{\text{eff},1}) - \frac{\alpha}{1+\alpha^2}[\boldsymbol{m_1} \times (\boldsymbol{m_1} \times \boldsymbol{h}_{\text{eff},1})] \\ \quad + \frac{\sigma_1 I_1 g_T}{1+\alpha^2}[\boldsymbol{m_1} \times (\boldsymbol{m_1} \times \boldsymbol{m_p}) - q(\boldsymbol{m_1} \times \boldsymbol{m_p})] \\ \frac{d\boldsymbol{m_2}}{d\tau} = -\frac{1}{1+\alpha^2}(\boldsymbol{m_2} \times \boldsymbol{h}_{\text{eff},2}) - \frac{\alpha}{1+\alpha^2}[\boldsymbol{m_2} \times (\boldsymbol{m_2} \times \boldsymbol{h}_{\text{eff},2})] \\ \quad + \frac{\sigma_2 I_2 g_T}{1+\alpha^2}[\boldsymbol{m_2} \times (\boldsymbol{m_2} \times \boldsymbol{m_p}) - q(\boldsymbol{m_2} \times \boldsymbol{m_p})] \end{cases} \quad (1)$$

Here α is the Gilbert damping coefficient, $\boldsymbol{m_i} = \boldsymbol{M_i}/M_{s,i}$ with $i = 1,2$ is the normalized magnetization vector of the free layer (FL) of the MTJ $i$ (1 = fixed-MTJ, 2 = free-MTJ) of saturation magnetization $M_{s,i}$ and volume $V_{FL,i}$, $\tau = \gamma_0 M_{s,i} t$ is the dimensionless time with $\gamma_0$ being the gyromagnetic ratio and $t$ the time. The effective field acting on the MTJ $i$ ($\boldsymbol{h}_{\text{eff},i}$) includes the contributions from the exchange ($\boldsymbol{h}_{\text{ex},i}$), uniaxial anisotropy ($\boldsymbol{h}_{u,i}$), demagnetization field ($\boldsymbol{h}_{\text{dmag},i}$) and the dipolar field ($\boldsymbol{h}_{\text{dip},j \to i}$) acting on MTJ $i$ due to MTJ $j$. The latter is the coupling mechanism of the two MTJs in the proposed device. The strength of the spin transfer torque (STT) is given by the term $I_i \sigma_i g_T$, where $I_i$ is the injected current in MTJ $i$ comprising both dc and ac component $I_i = I_{dc,i} + I_{ac,i} \sin(2\pi f_{ac} + \varphi_{ac})$, $\sigma_i = g\mu_B/(e\gamma_0 M_{s,i} V_{FL,i})$, and $g_T = 2\eta/(1 + \eta^2 \cos\vartheta)$ is the polarization function that depends on the angle formed between $\boldsymbol{m_i}$ and the pinned layer ($\boldsymbol{m_p}$) magnetization [39,40]. We observe that we consider, without loss of generality, that both MTJs have the same $g_T$ and $\boldsymbol{m_p}$. Micromagnetic calculations are performed by means of the micromagnetic solver PETASPIN utilizing an adaptive semi-implicit scheme [38,40]. In macrospin simulations, we exploit the uniform magnetization hypothesis to compute $\boldsymbol{h}_{\mathrm{dmag},i}$ as $\boldsymbol{h}_{\mathrm{dmag},i} = -\bar{\bar{N}} \boldsymbol{m}_i$ and $\boldsymbol{h}_{\mathrm{dip},j \to i}$ as $\boldsymbol{h}_{\mathrm{dip},j \to i} = (\mu_0/4\pi r_{ij}^3)[3\hat{\boldsymbol{r}}_{ij}(\boldsymbol{m}_j \cdot \hat{\boldsymbol{r}}_{ij}) - \boldsymbol{m}_j] = \bar{\bar{D}} \cdot \boldsymbol{m}_j$ via dipolar approximation, where $\bar{\bar{N}}$ is a diagonal matrix with the demagnetization factors as elements, $\hat{\boldsymbol{r}}_{ij}$ is the unit vector of magnitude $r_{ij}$ of the distance between MTJ $i$ and $j$, and $\bar{\bar{D}}$ is the position-dependent dipole matrix. All simulations are performed at zero temperature.

The spintronic MEMS accelerometer employs hybrid MTJs with an out-of-plane FL and in-plane PL patterned into pillars of elliptical cross section of dimensions 150 nm x 70 nm, designed according to previous schemes that optimize the rectification response of active spintronic diodes driven by the injection locking at zero bias field [41]. A sketch of the MTJs is also shown in Fig. 1(b). The FL is a 1.6 nm thick layer of $Co_{20}Fe_{60}B_{20}$, whereas the PL can be realized with a thicker ferromagnet or a synthetic antiferromagnet (SAF) exchange-biased to an antiferromagnet. The MTJ's magnetic parameters are saturation magnetization $M_s = 0.95$ MJ/(Tm$^3$), perpendicular magnetic anisotropy $K_u = 0.545$ MJ/m$^3$, exchange stiffness $A_{\mathrm{ex}} = 20$ pJ/m, Gilbert damping $\alpha = 0.02$, polarization efficiency $\eta = 0.66$, and a field-like torque that is 10% of the damping like torque, $q = 0.1$ [36,42]. The nominal electrical resistances associated with the parallel ($R_P$) and antiparallel ($R_{AP}$) configurations of the magnetization are 640 Ω and 1200 Ω, respectively [41]. Such MTJs have already been extensively studied experimentally and modelled by micromagnetic simulations [41,42].

To describe the mechanical dynamics of the MTJs, we exploit the fact that the device is designed to have the motion constrained along a fixed direction and that the time scales involved in elastic and magnetic processes are different. These allow us to restrict the dynamics to one dimension and to approximate the MTJ as a point-like object, whose position is given by its center of mass. Then, the elastic dynamics of the free-MTJ can be described as a one-dimensional driven damped spring-mass harmonic oscillator, governed by the following expression:

$$m_{\text{tot}} \frac{d^2 u_{\text{MTJ}}}{dt^2} = -b \frac{du_{\text{MTJ}}}{dt} - k_{el} u_{\text{MTJ}} - F_{\text{mag}} + F_{\text{ext}} . \quad (2)$$

The forces are applied to the center of mass of the free-MTJs and the free-MTJ's displacement ($u_{\text{MTJ}}$) is along the direction connecting the two MTJs, $m_{\text{tot}}$ is the total mass of the accelerometer system. The first term on the right hand side of the equation is the linear viscous damping with the viscous damping coefficient $b$, the second is the harmonic contribution with elastic constant $k_{el}$ and resonance frequency $\omega_0^2 = k_{el}/m_{\text{tot}}$, the third one is the magnetic interaction between the MTJs, and $F_{\text{ext}} = m_{\text{tot}} a_{\text{ext}}(t)$ is the driving external force acting on the device with external acceleration $a_{\text{ext}}(t)$. To calculate $F_{\text{mag}}$ in this one dimensional system we exploit the dipole approximation and the current-loop model [43]. This allows us to express the force acting on the magnetization of the FL of MTJ $i$ due to the stray field from MTJ $j$ ($\boldsymbol{h}_{\text{dip},j \to i}$) as $F_{\text{mag}} = \nabla(\boldsymbol{\mu}_i \cdot \boldsymbol{h}_{\text{dip},i \to j})$, where $\boldsymbol{\mu}_i = \boldsymbol{m}_i M_s V_{FL}$.

## IV. DESIGN OF THE FIXED-MTJ
### A. Magnetization dynamics of the fixed-MTJ

Since the MTJ working as an STNO is the main building block of the proposed device, we initially characterize its magnetization dynamics as a function of the dc current $I_{dc}$. We perform micromagnetic simulations and verify the uniform nature of the magnetization dynamics, necessary condition to be satisfied to apply a macrospin approximation. These are presented and discussed in Note S2 of Supplemental Material [32]. Fig. 3(a) shows the frequency $f$ and power $p$ (dimensionless unit) of the fixed-MTJ as a function of the dc current. Those data show the working region of the MTJ as STNO, the dynamic response is characterized by a negative nonlinear frequency shift $N/2\pi = df/dp$ around -0.75 GHz and a threshold current of -0.065 mA [44]. Once identified the dynamical properties of the STNO, we evaluate the rectification voltage in presence of an ac current $I_{ac} = 50\ \mu A$ with frequency $f_{ac} = 0.5$ GHz. The frequency is chosen to be close enough to the oscillation frequency of the oscillator in order to achieve the injection locking, and then an enhanced diode effect. Fig. 2(b) shows the rectified voltage $V_{dc}$ as a function of $I_{dc}$. The rectification voltage can be expressed analytically as $V_{dc} = (R_{AP} - R_P)\sqrt{p}/4 \ I_{ac} \cos[\varphi_{dc}(I_{dc})]$ following Ref. [42]. The maximum $V_{dc}$ ($V_{dc,max}$) is obtained for $\varphi_{dc}(I_{dc}) = 0$. By fitting the quadratic expression of $V_{dc}$ as a function of $I_{dc}$, (solid line in Fig. 2(b)), we extract the dc current that gives $V_{dc,max}$ to be $I_{dc} = -0.09$ mA. Similar qualitative results have been observed for $I_{ac} = 25 - 75 \ \mu A$. The validity of the $V_{dc}$ expression has been checked with numerical calculations of the oscillator power $p$ ($\sqrt{p} = dm_x$) and the intrinsic phase shift ($\varphi_{dc}$) between the ac current and the oscillating magnetization as a function of the $I_{dc}$, as summarized in Fig. 3(c). Here it can be clearly observed the linear dependence of the intrinsic phase shift ($\varphi_{dc}$) and weak dependence of $dm_x$ ($dm_x$ variation is less than 1 %), as expected for STNO with a large nonlinear frequency shift [44]. Fig. 2(d) shows the time traces of the input ac signal and oscillating MTJ magnetoresistance ($\Delta R$) for $I_{dc} = -0.09$ mA. $\Delta R$ is computed considering the following expression: $\Delta R^{-1} = \frac{(R_P^{-1}+R_{AP}^{-1})}{2} + \frac{(R_P^{-1}-R_{AP}^{-1})}{2}\cos(\beta)$ [41], where β is the angle between the FL magnetization $\boldsymbol{m}$ and $\boldsymbol{m_p}$. We point out that since the polarizer is magnetized along the x-direction, the oscillations in $\Delta R$ is due to $dm_x$. For this reason, $\varphi_{dc}(I_{dc})$ of the $\Delta R$ is the same of the $dm_x$. These results agree with previous micromagnetic simulations and experimental reports [41,42], and represent the fundamental brick of our system.

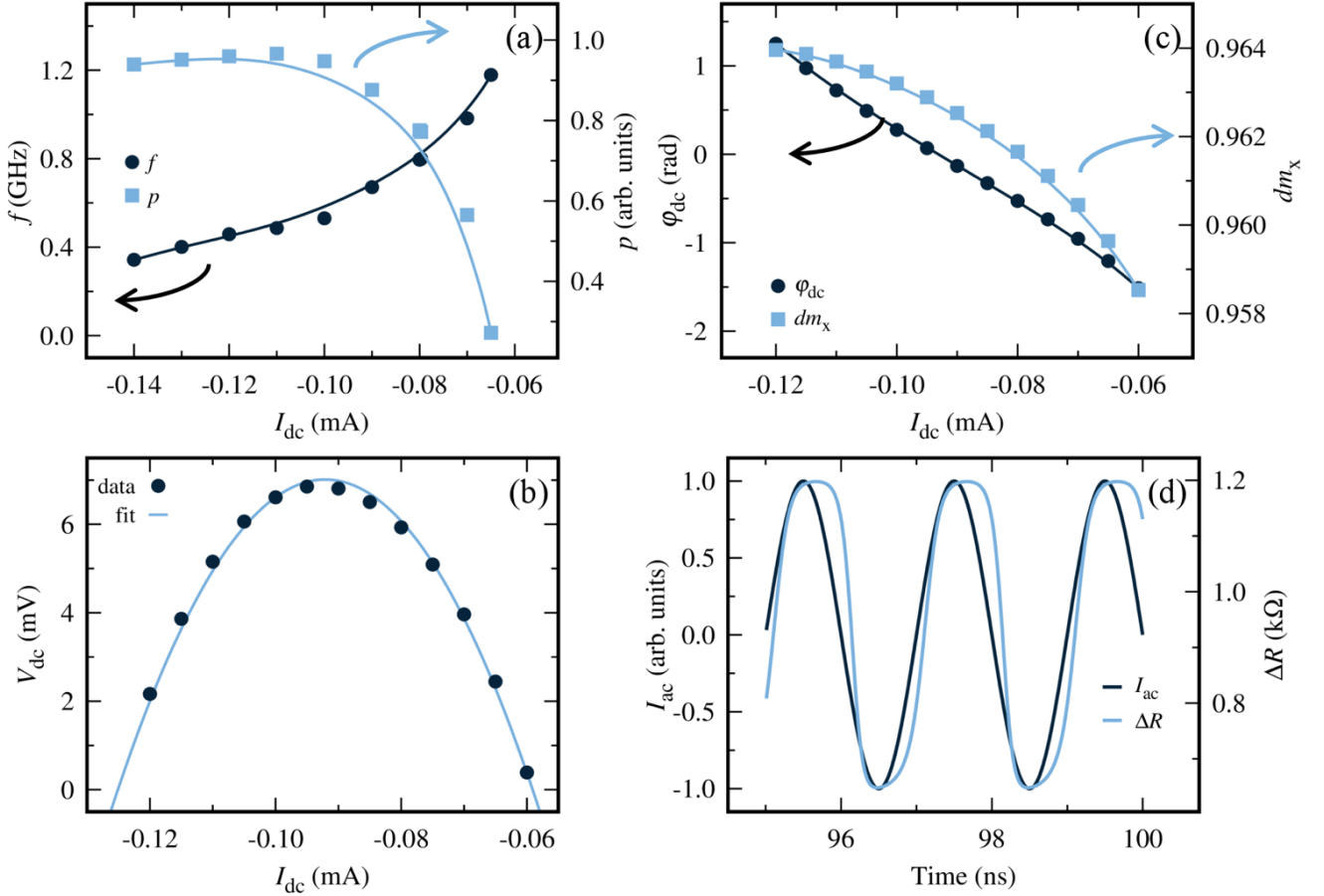

Fig. 2 (a) Oscillation frequency ($f$) and power ($p$) of the x-component of the FL magnetization as a function of the applied dc current ($I_{dc}$), lines are guides to the eye. (b) Rectification voltage ($V_{dc}$) as a function of $I_{dc}$ (symbols) for $I_{ac} = 50\ \mu A$ and $f_{ac} = 0.5$ GHz, and fit (solid line) according to $V_{dc} = (R_{AP} - R_P)\sqrt{p}/4\ I_{ac,max} \cos[\varphi_{dc}(I_{dc})]$. (c) intrinsic phase shift ($\varphi_{dc}$) and amplitude of the x-component of the magnetization ($dm_x$) as a function of $I_{dc}$ for a fixed ac current $I_{ac} = 50\ \mu A$ at a frequency $f_{ac} = 0.5$ GHz. Lines are guides to the eye. (d) A comparison of time traces of the input current $I_{ac}$ and the oscillating magnetoresistance ($\Delta R$) for $I_{dc} = -0.09$ mA, $I_{ac} = 50\ \mu A$, and $f_{ac} = 0.5$ GHz.

## V. MUTUAL SYNCHRONIZATION OF THE FREE- and FIXED- MTJs

Once the single STNO has been characterized, as the first step towards modeling the accelerometer device, we studied the dynamics of the synchronized state of the two MTJs at different distances ($d$ = center-to-center distance, see Fig. 1(b)). We set the fixed-MTJ into the injection-locked state upon application of ac ($I_{ac}$) and dc ($I_{dc,1}$) currents. The free-MTJ is then biased with a second dc current ($I_{dc,2}$) which also drives self-oscillations. The two MTJs are coupled via the stray field generated by the two free layers (we consider the pinned layer as an exchange biased SAF, which generates a reduced bias dipolar field). To compute the stray field, we apply the dipole approximation and we consider $d > 350$ nm. At each position we compute the position-dependent dipole matrix ($\bar{\bar{D}}$) and calculate its product with the other MTJ magnetization vector $\boldsymbol{m}$. When the two MTJs are far apart, the magnetic interaction is negligible, and their magnetization dynamics are independent. As the MTJs get closer, the dipolar interaction couples the two MTJ dynamics, leading to a mutual synchronization [45]. We wish to stress that finding such conditions is crucial for the operation of the spintronic MEMS accelerometer. Based on the characterization of the single STNO, to set the injection-locking in the fixed-MTJ we apply $I_{dc,1} = -0.09$ mA, $I_{ac} = 50$ µA, and $f_{ac}=0.5$ GHz. Then, we perform simulations as a function of the center-to-center distance between the MTJs ($d$), highlighted in Fig. 1(b), by varying $I_{dc,2}$ for values around $I_{dc,1}$. The rectification voltage ($V_{dc,0}$) of the fixed-MTJ when it is isolated is used as a reference for the calibration of the system. Hence, the rectification voltage can be expressed as $\Delta V_{dc} = V_{dc} - V_{dc,0}$, where $V_{dc}$ is the rectification voltage measured across the fixed-MTJ when the two MTJs are coupled. Fig. 3(a) shows the obtained $\Delta V_{dc}$ as a function of $d$ in the range 350 nm to 650 nm for various $I_{dc,2}$ (see figure caption). For the parameters we are considering here, the mutual synchronization occurs for $I_{dc,2}$ close to $I_{dc,1}$. In the rest of the work, we will consider $I_{dc,2} = -0.10035$ mA (see Fig. 3(b)), but similar qualitative results are achieved for other curves.

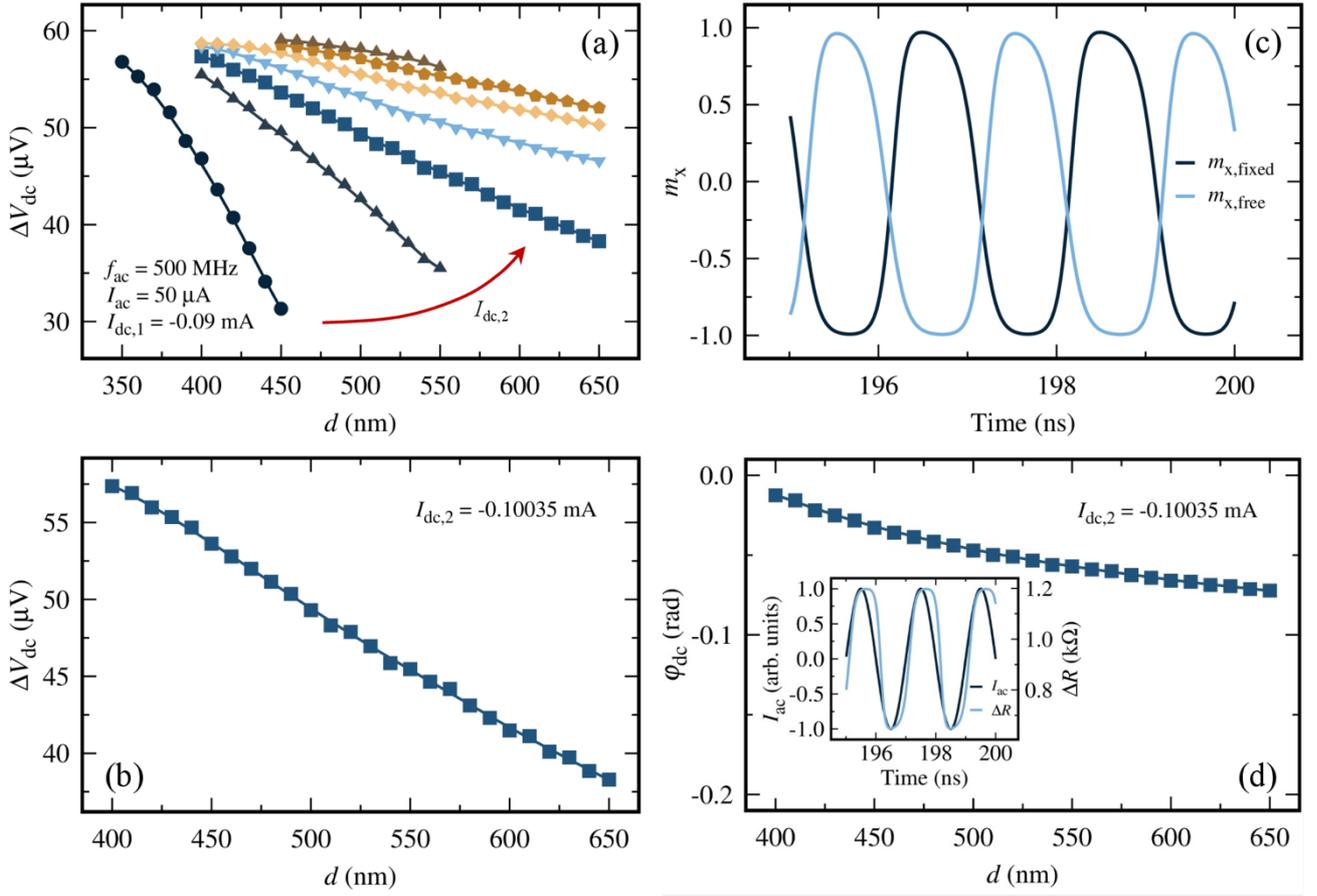

Fig. 3 (a) Rectification voltage $\Delta V_{dc}$ as a function of the distance between the two MTJs for different dc currents injected into the free-MTJ $I_{dc,2}$ (-0.099 mA black dots, -0.1 mA upwards dark blue triangles, -0.10035 mA blue navy squares, -0.1008 mA downwards light blue triangles, -0.10135 mA yellow diamonds, -0.1017 mA light brown pentagons, 0.10225 mA upward dark brown triangles) and (b) focus on the case $I_{dc,2} = -0.10035$ mA. (c) Plot of the fixed- and free-MTJs $m_x$ components at a distance $d = 450$ nm for $I_{dc,2} = -0.10035$. (d) Intrinsic phase shift ($\varphi_{dc}$) as a function of $d$ for $I_{dc,2} = -0.10035$ mA. The inset shows the time traces of the input ac signal $I_{ac}$ and magnetoresistance $\Delta R$ for $I_{dc,2} = -0.10035$ mA and distance $d = 450$ nm.

Fig. 3(c) shows an example of the time evolution of $m_x$ for the fixed- and free-MTJs upon application of $I_{dc,1} = -0.09$ mA, $I_{ac} = 50$ μA, $f_{ac} = 0.5$ GHz and $I_{dc,2} = -0.10035$ mA at $d = 450$ nm where the phase difference is near 180°. However, it is worth pointing out that for a given set of $I_{dc,1}$, $I_{ac}$ and $I_{dc,2}$, the phase difference depends on the distance between the MTJs, and it is determined by the dipolar coupling. Therefore, it may be adjusted by tuning these parameters. The inset of Fig. 3(d) compares an example of time domain traces of magnetoresistance, induced by the oscillations of $m_x$ of the fixed-MTJ, and the input ac signal for a distance $d$ = 450 nm. The phase difference between the two signals varies as $d$ changes (see main panel of Fig. 3(d)), while the power remains constant (not shown). Hence, it is the distance dependent phase shift that is the key ingredient for the change of $\Delta V_{dc}$ as a function of the distance $d$. This is advantageous as the phase offers a robust tool, and it can be easily accessed by measurements.

In Note S3 of Supplemental Material [32] we discuss micromagnetic simulations performed to verify the effect that non-uniformities in the stray field distribution, due to the quasi-uniform magnetization of the MTJs, may have on the magnetization dynamics. Despite the slightly higher frequencies at which the synchronization occurs, we do not observe qualitative difference with the macrospin results.

## VI. SPINTRONIC MEMS ACCELEROMETER
### A. Results

In modeling the whole magnetic-elastic dynamics, we can consider a range of realistic situations such as car crash, screen re-orientation, and seismic events. Those excitations are generally characterized by a low frequency response ranging from few Hz to kHz. For this reason, usually, capacitive MEMS-based accelerometers [12,46,47] are designed to have mechanical resonance frequencies ($f_0 = \omega_0/2\pi$) between 1 and 100 kHz, which allows to consider the quasi-static response of the oscillator in the process of extracting the acceleration. In such a condition, when the mechanical resonance is at least 10 times larger than the external acceleration, the latter can be determined from the equilibrium condition of the forced harmonic oscillator $a_{ext} = \omega_0^2 ds$, where $ds$ is the displacement. Considering the same substrate technology of current MEMS-based devices, the spintronic MEMS accelerometer can exploit the same approach to extract the acceleration directly from the displacement, in our case $d$. In particular here, we assume a total mass of the system $m_{tot} = 1$ µg and an elastic constant of the spring $k_{el} = 1$ KN/m [12,46,47]. This combination of $m_{tot}$ and $k_{el}$ gives a resonance frequency of the system $f_0 \sim 5$ kHz that is order of magnitudes lower than the ferromagnetic resonance frequency (GHz range); we also consider a typical damping ratio $\zeta = \frac{b}{\sqrt{4k_{el}m_{tot}}} = 0.75$. With such elastic properties, the distance between the MTJs at equilibrium is 500 nm. This value corresponds to the midpoint of the range in which $\Delta V_{dc}$ exhibits a linear dependence with $d$, as shown previously in Fig. 3(b).

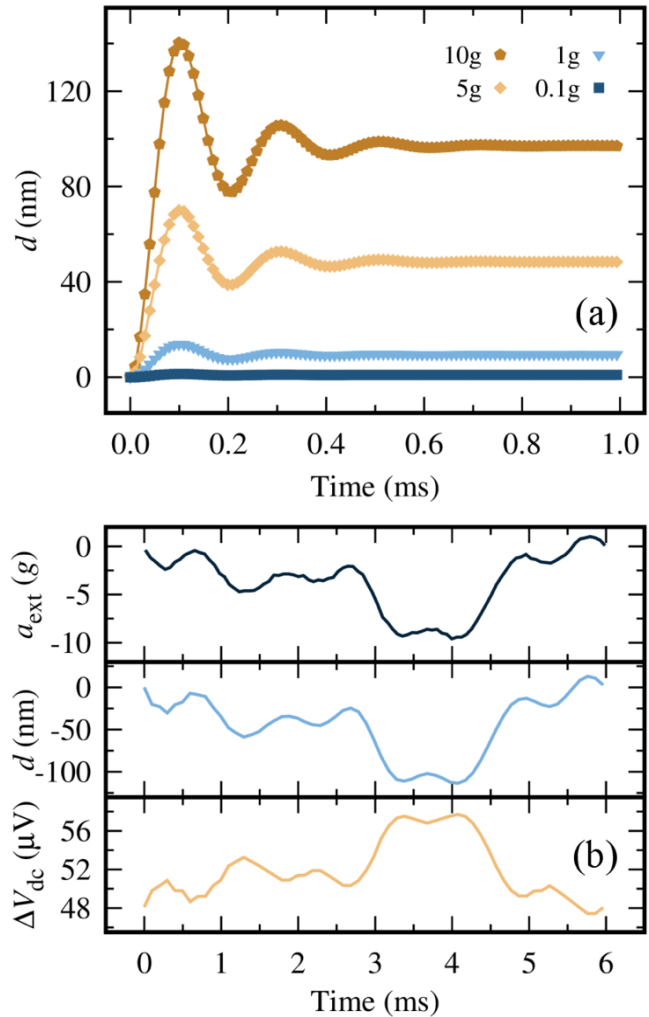

Fig. 4 (a) Displacement of the free-MTJ with respect to its initial position as a function of time when the device is subjected to different constant external accelerations, given in units of $g$. (b) Input time-dependent external acceleration $a_{\text{ext}}$ given in units of acceleration of gravity $g$ (top panel), relative displacement $d$ of the free-MTJ from the fixed-MTJ (middle panel) and time-dependent rectification voltage $\Delta V_{dc}$ measured at the fixed-MTJ (bottom panel) in response to $a_{\text{ext}}$.

As an example, let us set the fixed-MTJ in injection locking regime by applying $I_{\text{dc},1} = -0.09$ mA, $I_{\text{ac}} = 50$ µA and $f_{\text{ac}} = 0.5$ GHz. We also inject a current $I_{dc,2} = -0.10035$ mA to generate the self-oscillation in the free-MTJ. Fig. 4(a) shows the time evolution of the distance $d$ for different constant accelerations up to $10g$, where $g = 9.81$ m/s² is the acceleration of gravity. The transient elastic dynamics elapses in less than 0.5 ms, and the final displacement $d$ corresponds to the forced harmonic case, $a_{\text{ext}} = \omega_0^2 d$. In addition, we find that the transient of the magnetization dynamics occurs in less than 40 ns (not shown here). This time interval is negligible when compared to the elastic dynamics, and it does not impose limitations to the operability of the proposed device. This direct mapping between acceleration and voltage potentially allows one to simplify the acquisition process, the extra circuitry and the post-processing to evaluate the acceleration. In other words, to determine the displacement, we exploit its linear effect on the output voltage $\Delta V_{dc}$, demonstrated in the static case shown in Fig. 3(b) and confirmed from Fig. 4(a). By performing a preliminary calibration of the response of $\Delta V_{dc}$ as a function of $d$, it is possible to translate the measured output voltage into the displacement, and consequently the acceleration $a_{\text{ext}}$.

Once the device has been validated for constant accelerations, we consider a more realistic case by mimicking the acceleration and deceleration profile of a car, for instance to decide whether to issue an attention warning to the driver. The top panel of from Fig. 4(b) shows the $a_{\text{ext}}(t)$ that we provide as input to the accelerometer. We employ the same mechanical parameters utilized previously and simulate the response of the system under the external acceleration $a_{\text{ext}}(t)$, which has a maximum amplitude of $-10g$. With these parameters, the mechanical oscillator adapts to the external acceleration within 0.5 ms, thus few ms are sufficient to evaluate the performance of the system. The middle and bottom panel of Fig. 4(b) show the time evolution of the relative displacement $d$ of the free-MTJ with respect to the fixed-MTJ as well as the time trace of $\Delta V_{dc}$ generated across the fixed-MTJ in response to $a_{\text{ext}}(t)$, respectively. From the figure, the linear relationship between $d$ and $\Delta V_{dc}$ is evident by inspection: the variation of $\Delta V_{dc}$ follows that of $d$, with the two curves overlapping perfectly once the sign of $\Delta V_{dc}$ is changed. In fact, for negative displacements (free-MTJ moving closer to the fixed-MTJ) $\Delta V_{dc}$ is positive, as clear from Fig. 3(b). We record a variation of $\Delta V_{dc}$ of the order of 10 µV, as expected from the static case and the constant acceleration test. The comparison of the three panels of Fig. 4(b) shows clearly the direct mapping between $\Delta V_{dc}$, $d$ and $a_{\text{ext}}$, proving the goodness of the proposed approach. We would like to stress that the simulated timescale is faster than the actual event, in which $a_{\text{ext}}$ varies in the range of seconds or longer. The choice has been dictated by the computational time required to access such long times, and the results do not exhibit any effect attributable to an excessively fast dynamics. On the contrary, it is clear that in the case of external accelerations with a lower frequency, the proposed spintronic MEMS device would operate successfully.

In the absence of a whole circuitry evaluation, to estimate the energy consumption of the device due to the injection of dc currents in both the MTJs, required to exploit the spintronic features of the MTJs, one can extract it from the average MTJ resistance ($R_{MTJ,\text{avg}}$). Following the work of Zeng *et al.* [48] on analogous MTJ devices (refer to the current region marked as B1 in Figure 1.c), $R_{MTJ,\text{avg}}$ can be estimated around 800 Ω. Thus, by considering working currents $I_{dc,\{1,2\}}$ of -0.1 mA, we can estimate a power consumption lower than 20 µW taking into account both the free- and fixed-MTJs. This estimated power consumption is comparable with or better than those of commercial MEMS-based devices, as reported in data sheets of commercially available accelerometers.

### B. Sensitivity of the spintronic MEMS accelerometer

There exist various figures of merit that characterize a sensor depending on the transduction mechanisms involved [12,47,49]. One of the most common in capacitive-MEMS-based accelerometers is the capacitance sensitivity, i.e., the change in capacitance with respect to the external acceleration. However, the absence of the capacitive element, which is a specificity of the proposed idea, does not allow us to make a comparison to this figure of merit. On the other hand, we can use a sensitivity defined as the ratio between the amplitude of the external acceleration and the read-out voltage. Since the variation of $\Delta V_{dc}$ is on the order of tens of microvolts, as shown in Fig. 3 and Fig. 4, we obtain sensitivities on the order of 0.5 µV/$g$ considering directly the voltage extracted across the MTJ without any amplification or postprocessing. We wish to stress that the aim of this work is to present a novel design that may drive the development of a new type of devices based on spintronics. Optimization of the device has not been performed at this stage. However, by tuning opportunely the MTJs setup properties higher output voltages and higher sensitivities with the spintronic diode effect can be obtained. The displacement or mechanical sensitivity, proportional to $1/\omega_0^2$, relates to the mechanical response of the device and in capacitive MEMS-based accelerometers is related to the displacement of the proof mass along the measurement axis and external acceleration. In our design, we can relate the displacement of the free-MTJ with the acceleration by applying different accelerations and extract it as the slope of the linear relationship between the two quantities. Considering the same substrate technology, we obtain a mechanical sensitivity around 10 nm/$g$ for the chosen mechanical parameters.

## VII.    SUMMARY AND CONCLUSIONS

We have proposed a spintronic MEMS accelerometer based on two MTJs magnetically and elastically coupled both working as STNO with bias currents on the order of the mA. The oscillatory frequency of the fixed-MTJ is controlled by an ac current that drives the injection locking to enable the use of the spintronic diode effect as the read-out mechanism of the acceleration applied to the device. We emphasize that the elastic connection between the two substrates on which the MTJs are fabricated can also be realized exploiting the same technology used for capacitive MEMS accelerometers, which makes the proposed device technologically appealing. Moreover, while the intrinsic magnetic coupling between the MTJs is often considered disruptive in applications [50–52], here it plays a crucial role due to its strong dependence on the position. The idea of spintronic MEMS accelerometer can be scalable to smaller system and the sensitivity can be enhanced by designing a network of spintronic MEMS. For example, another possible implementation would consist of three aligned MTJs where the fixed-MTJ is placed in between two free-MTJs. When the MEMS is subjected to the acceleration, one of the free-MTJs is displaced further away from the fixed-MTJ, the other closer. This solution would offer an improvement in both the sensitivity and the reliability of the device.

Overall, the major potential advantages of the spintronic MEMS accelerometer are the compatibility with current technological implementations of MTJs, a robust and simple reading scheme [20,30], other than the low power requirements. In fact, the spintronic accelerometer proposed here senses the acceleration directly as voltage variation across the fixed-MTJ in contrast to state of the art integrable MEMS-based devices where a circuit to convert the capacitance variation into an output voltage is needed. In conclusion, we believe that this idea can be of stimulus for the spintronic community to design new experiments and expand the application of MTJs for sensors.

**ACKNOWLEDGEMENTS**

The research has been supported by project no. PRIN 2020LWPKH7 funded by the Italian Ministry of University and Research and by Petaspin association (https://www.petaspin.com/). The work at the University of Messina has been also supported under the project number 101070287 — SWAN-on-chip — HORIZON-CL4-2021-DIGITAL-EMERGING-01 funded by the European Union. DR acknowledges the project PON Ricerca e Innovazione D.M. 10/08/2021 n. 1062. The work at Northwestern University was supported by the National Science Foundation (NSF), Division of Electrical, Communications and Cyber Systems (ECCS), under award number 2203242.

SUPPLEMENTAL MATERIAL

A magneto-mechanical accelerometer based on magnetic tunnel junctions

Andrea Meo[1], Francesca Garescì[2], Victor Lopez-Dominguez[3-4], Davi Rodrigues[1], Eleonora Raimondo[5], Vito Puliafito[1], Pedram Khalili Amiri[4,*], Mario Carpentieri[1,*] Giovanni Finocchio[5,*]

[1] Department of Electrical and Information Engineering, Politecnico of Bari, 70125 Bari, Italy

[2] Department of Engineering, University of Messina, I-98166, Messina, Italy

[3] Institute of Advanced Materials (INAM), Universitat Jaume I, Castellon, 12006, Spain

[4] Department of Electrical and Computer Engineering, Northwestern University, 2145 Sheridan Road, Evanston, 60208, Illinois, USA

[5] Department of Mathematical and Computer Sciences, Physical Sciences and Earth Sciences, University of Messina, I-98166, Messina, Italy

*Corresponding authors: pedram@northwestern.edu, mario.carpentieri@poliba.it, gfinocchio@unime.it

# Supplemental Note 1 - Alternative working principle

An alternative spintronic MEMS design, similar to the one proposed in Fig. 1(b) of the main manuscript can be proposed. In this alternative scheme, presented in Fig. S1, a coil is placed in between the MTJs placed on top of two substrates connected by means of elastic elements. The coil allows to electrically read out the stray field changes induced by the acceleration, corresponding to a displacement of the free-MTJ. The variation of the stray field upon application of an external acceleration induces an electromotive force across the coil terminals that can be directly linked to the acceleration. In such a device the MTJ are utilized as passive magnets whose advantages are the nanometric dimensions, the possibility of controlling and resetting the magnetic state of the MTJs free layers and the non-volatility. The spintronic MEMS considered in the main manuscript (shown in Fig. 1(b)) presents main advantages with respect to the solution shown here: it does not require the extra coil component and the detection is performed exploiting the spintronic diode effect. Those features make such a design potentially more scalable, reliable and sensitive than the one discuss here, that can instead be considered as a first and more simplistic attempt to a spintronic MEMS-based accelerometer.

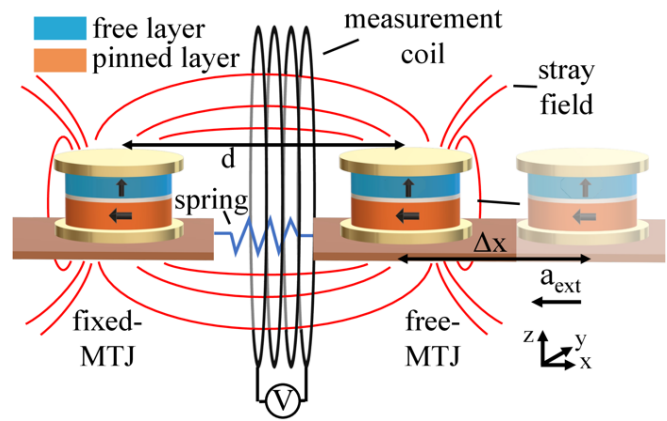

Fig. S1: Schematic description of a spintronic MEMS accelerometers. A system of two MTJs, composed of free layer and pinned layer, is coupled via dipolar field (red lines) and elastically via the substrates. The free-MTJ is displaced from its equilibrium position (shaded area) of Δx when an external acceleration $a_{ext}$ is applied, resulting in a change in the MTJs center-to-center distance (d) and thus in the stray field distribution. The read out mechanism exploits the change in the induced voltage in the measurement coil. The Cartesian reference system is also included.

## Supplemental Note 2 - Evaluation of the macrospin hypothesis

We perform micromagnetic simulations to verify the validity and applicability of the macrospin approximation. In micromagnetic simulations the FL is discretized in cells of dimensions 2.5 nm x 2.5 nm x1.6 nm. We fist characterize the MTJ magnetization dynamics as a function of the dc current $I_{dc}$. The device is designed to have the major oscillation in the x-component of the magnetization ($m_x$), as shown below. Fig. S2(a) shows an example of comparison for the time domain dynamics of the magnetization for $I_{dc} = -0.1$ mA, light blue and black lines are with full micromagnetic simulations and within the macrospin approximation respectively. From this comparison, it can be clearly seen a good agreement. Fig. S2(b) shows a significative example of the snapshot of the magnetization (t = 1.2 ns red dot on Fig. S2(a)) which exhibits a quasi-uniform magnetization distribution with only some tilting away at the edges of the ellipse. By comparing the average magnetization length ($\langle m_{l,\text{micro}} \rangle$) with the value expected for a uniform system, i.e., $1 - \langle m_{l,\text{micro}} \rangle$, we find that the micromagnetic results are characterized by a degree of non-uniformity of less than 2%. Since micromagnetic calculations become time consuming and computationally expensive when the time interval to be simulated is about hundreds of nanoseconds or more, we can lift some of the burden by utilizing the macrospin approach. Thus, macrospin simulations will be utilized in the following unless otherwise specified.

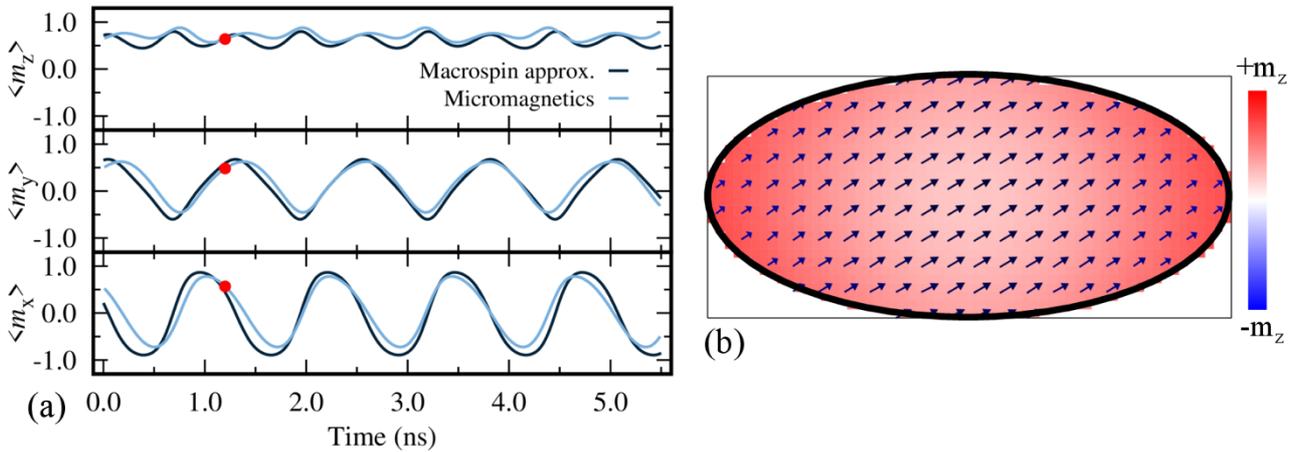

Fig. S2: (a) Comparison between macrospin (black) and micromagnetic (light blue) dynamics of the magnetization for the same system. (b) Snapshot of the magnetization of the STNO FL captured at 1.2 ns (indicated by the red dots in panel a)) showing a quasi-uniform configuration. Arrows represent the in-plane component of the magnetization, the color bar describes the normalized z-component of the magnetization $m_z$ (red=+z, white=in-plane and blue=-z).

# Supplemental Note 3 - Micromagnetic verification of mutual synchronization between the MTJs

The essential ingredient of the proposed spintronic MEMS is the magnetic coupling between the MTJs. To further verify the validity of the macrospin approach, we perform micromagnetic simulations to include the non-uniform stray field generated by the neighbor MTJ. We place the MTJs at a center-to-center distance $d = 400$ nm and synchronize the MTJs dynamics injecting a dc current $I_{dc,1} = -0.1$ mA and ac current of amplitude $I_{ac} = 50$ µA and frequency $f_{ac} = 0.8$ GHz in the fixed-MTJ working as STD, and a dc current $I_{dc,2} = -0.12$ mA in the free-MTJ. Fig. S3(a) shows the time traces of the average components of the magnetization $\langle m_{x,y,z} \rangle$ for both the MTJs, where we can observe the synchronization of the two dynamics. There is a small difference in the current magnitudes with respect to the macrospin case discussed in the main manuscript, that we ascribe to the small non-uniformities in the micromagnetic magnetic texture. These small non-uniformities are also responsible for the higher input frequency at which the synchronization occurs in the micromagnetic case. In panels b), c) and d) we show snapshots of the magnetic configuration of the two MTJs throughout the time evolution. The quasi-uniform state is characterized by a magnetization that tends more towards an in-plane configuration at the center of the ellipse than at the edges along the long-axis of the ellipse, visible as a whiter region (color palette in the snapshots represents $m_z$), more pronounced in the fixed-MTJ. The latter can be understood as the system is driven close to resonance by the simultaneous injection of dc and ac currents and thus to a stronger excitation, whereas the former can be expected since the demagnetizing contribution is stronger at the center of the disc, as observed in section S3. However, it is worthy pointing out that we do not observe formation of domains or non-collinear magnetic structures during the oscillatory dynamics in any of the MTJs and that the magnetization is overall uniform. Fig. S3(e) shows the degree of non-uniformity, defined as $1 - \langle m_l \rangle$ as in the previous section, from which we can observe that the micromagnetic magnetization dynamics is characterized by a degree of non-uniformity of less than 3% and 2% for the fixed-MTJ and free-MTJ, respectively. Moreover, the fact that the magnetic system may oscillate at higher frequencies would prove beneficial for the device, as this is one of the main condition required to fully exploit the potential of the proposed spintronic MEMS accelerometer. Thus, these results confirm that the system synchronization occurs and is robust even in the presence of a non-uniform stray field distribution, and they allow for a macrospin approximation even though taking into account possible differences with a full micromagnetic description.

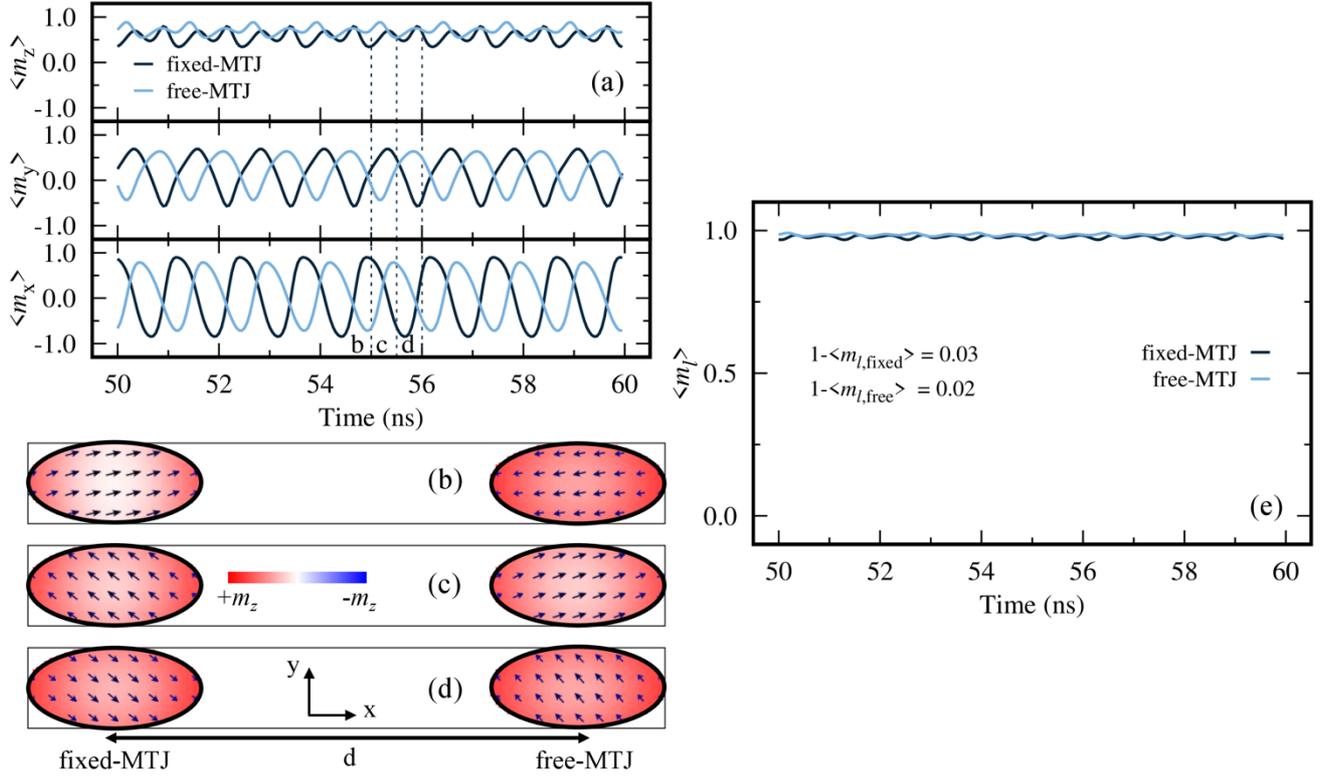

Fig. S3: (a) Comparison between fixed-MTJ (black) and free-MTJ (light blue) magnetization dynamics when the MTJs are at a center-to-center distance $d = 400$ nm for $I_{dc,1} = -0.1$ mA, $I_{ac} = 50$ µA, $f_{ac} = 0.8$ GHz and $I_{dc,2} = -0.12$ mA. (b,c,d) Snapshots of the magnetization of the FL magnetization corresponding to point b, c, d indicated by dashed lines in a); the color bar describes the normalized z-component of the magnetization $m_z$ (red=+z, white=in-plane and blue=-z) and we also indicate the Cartesian reference system. e) Plot of the degree of non-uniformity $(1 - \langle m_l \rangle)$ corresponding to the magnetization dynamics plotted in a), where $\langle m_l \rangle$ is the average magnetization length.